\documentclass{aastex63}
\usepackage{amsmath}
\usepackage{amssymb}
\usepackage{graphics}
\usepackage{xcolor}
\usepackage{lineno}
\usepackage{hyperref}
\usepackage{booktabs}
\usepackage[caption = false]{subfig}
\usepackage{float}

\received{}
\revised{}
\accepted{}
\submitjournal{ApJ}

\shorttitle{Effect of vegetation on {Trappist-1}}
\shortauthors{Vecchio et al.}

\begin{document}

\title{Effect of vegetation on the temperatures of {Trappist-1} planets}

\correspondingauthor{Antonio Vecchio}
\email{a.vecchio@astro.ru.nl, antonio.vecchio@obspm.fr}
\author[0000-0002-2002-1701]{Antonio Vecchio}
\affiliation{Radboud Radio Lab, Department of Astrophysics/IMAPP-Radboud University, P.O. Box 9010,
6500GL Nijmegen, The Netherlands}
\affiliation{LESIA, Observatoire de Paris, Universit\'e PSL, CNRS, Sorbonne Universit\'e, Universit\'e de Paris, 5 place Jules Janssen, 92195 Meudon, France}

\author{Leonardo Primavera}
\affiliation{Dipartimento di Fisica, Universit\`a della Calabria, Ponte P. Bucci, Cubo 31C, 87036, Rende (CS), Italy}

\author{Fabio Lepreti}
\affiliation{Dipartimento di Fisica, Universit\`a della Calabria, Ponte P. Bucci, Cubo 31C, 87036, Rende (CS), Italy}

\author{Tommaso Alberti}
\affiliation{INAF-Istituto di Astrofisica e Planetologia Spaziali, via del Fosso del Cavaliere 100, Roma, Italy}

\author{Vincenzo Carbone}
\affiliation{Dipartimento di Fisica, Universit\`a della Calabria, Ponte P. Bucci, Cubo 31C, 87036, Rende (CS), Italy}


\begin{abstract}
TRAPPIST-1 is an ultra-cool dwarf hosting a system consisting of seven planets. While orbital properties, radii and masses of the planets are nowadays well constrained, one of the open fascinating issues is the possibility that an environment hospitable to life could develop on some of these planets. Here we use a simple formulation of an energy balance model that includes the vegetation coverage to investigate the possibility of life affecting the climate of the planets in the TRAPPIST-1 system. Results confirm that planet T-e has the best chance for a habitable world and indicate that vegetation coverage significantly affects the resulting temperatures and habitability properties. The influence of vegetation has been evaluated in different scenarios characterized by different vegetation types, land-sea distributions and levels of greenhouse effect. While changes in vegetation type produce small changes, about ${0.1\%}$, in the {habitable surface fraction}, different land-sea distributions, by also affecting the vegetation growth, produce different temperature distributions. Finally  at latitudes where vegetation grows, the lowering of local albedo {still represents a relevant contribution in settling the planetary temperature profiles even when  levels of greenhouse effect higher than the Earth-like case are considered.}
\end{abstract}



\section{Introduction}
{ The large number of exoplanets discovered in recent years \citep{Mayor95,Marcy96,Petigura13,Gillon16,Gillon17} has led to a remarkable increase of  scientific interest in this topic, as also confirmed by the number of dedicated missions developed by NASA (\url{exoplanets.nasa.gov/exep/}) and ESA  (\url{sci.esa.int/home/51459-missions/}).
One of the open issues about exoplanets, maybe the most fascinating one, concerns their possible habitability. The very limited amount of information about physical properties of planets and atmospheres is mainly inferred through spectroscopic measurements \citep[e.g.][]{dewit18} and estimations of composition is made by mass-radius fitting through dynamical models \citep{quarles17,uinterborn18}.  Climate models, that use this information as input, can be useful to gain insight into the possible habitability. Different approaches, from simple formulations of energy balance models (EBMs)  \citep[see e.g][]{Pierrehumbert11, Alberti17} to more elaborate three-dimensional climate models \citep[see e.g][and references therein]{Gillon17,Wolf17}, where used to investigate this topic. 

{The Trappist-1 planetary system is formed by seven planets orbiting around an ultracool dwarf  hosting star \citep{Gillon16,Gillon17,Luger17,Kislyakova17}. While initially all the planets seemed to be consistent with an Earth like composition \citep{quarles17}, recent analysis of transit timing variations show that Trappist-1 c and e likely have largely rocky interiors, while  other planets require volatile-rich envelopes  in most cases with water mass fractions less than $5\%$ \citep{Grimm18}.} The {seven} planets are enclosed within 0.062 AU from the star but the stellar luminosity $S_\star$ is fainter with respect to our Sun (L$_\star \sim 5.25 \times 10^{-4} L_\odot$, the symbol $\odot$ refers to solar parameters at 1 AU) and the net flux $S_\star$ impinging on the planets is
in the range $0.13 S_\odot \le S_\star \le 4.3 S_\odot$ \citep{Gillon16,Gillon17,Wolf17}.  According to \citet{Gillon16}, the seven {Trappist-1} planets are temperate, having equilibrium temperatures below $\sim400$ K. 
Due to the proximity to the hosting star, the planets are likely to synchronously rotate with it, thus generating a permanent dark side on the planetary surface \citep{Joshi97,Heath99,Kopparapu14,Kopparapu16}. This has a strong impact on the planetary climate since on tidally locked planets the tidal synchronization produces heat redistribution by means of atmosphere large-scale heat transport through fluid motions, driven by the day-night temperature contrast \citep{Koll16, Komacek16,Joshi03,Merlis10,Forget14,Wordsworth15,Joshi97,Pierrehumbert11}. \\
Concerning the habitability of the planets belonging to the {Trappist-1} system, several approaches have been used to investigate the possibility that some of them can reside in the so-called circumstellar Habitable Zone (HZ), the area around a star where a planet can sustain liquid water on its surface \citep{Kasting93,Kopparapu13}.
By using radiative-convective climate models, that resolve the fluid dynamics of the atmosphere under different atmospheric compositions (mostly N$_2$, CO$_2$, and H$_2$O) and assuming completely ocean-covered planets, it was found that planet T-e has the most favorable conditions for habitability, T-d presents a runaway greenhouse effect, while inner (outer) planets have higher (lower) surface temperatures than the upper (lower) limit for residing in the HZ  \citep{Wolf17,Bolmont17,Bourrier17,Kopparapu17}. 
Results from a 0-D Daisyworld-like energy-balance model show that planets d and e are the most likely to be habitable \citep{Alberti17,Barr18}.

Here we take into account the possibility of life affecting the planetary climate by developing a 1-D EBM that includes an evolving vegetation coverage, acting as a feedback on the surface temperature of the tidally locked planets of the {Trappist-1} system. The model describes the evolution of the surface temperature  $T(\theta,t)$ and fraction of vegetation coverage $A(\theta,t)$, both functions of time and { the tidally locked} latitude $\theta$. The main advantage of using an EBM with respect to global climate models, although the latter provide a more complete description, is that it allows to explore a wider parameter space, reduces the large pool of assumptions of atmospheric and surface conditions required by more complex climate models, and provides direct information about global properties of the planets such as their habitability. At the same time, global models are less efficient in evaluating the effects of vegetation on the planetary temperature and in resolving the physics behind vegetation feedbacks. The possibility to investigate the effects of vegetation represents a peculiarity of this model and distinguishes it from other works studying the climate of terrestrial exoplanets that do not take into account the feedback of life on the climate. 
In discussing the results of our model, we show that including the vegetation coverage in exoplanet climate modelling is crucial since it can produce significant features in the climate and affect the habitability.  

\section{The model} \label{sec:model}
{The equations of the model only depend on the tidally locked latitude $\theta$; variations along the longitudinal direction are neglected as in similar models  \citep{Pierrehumbert11, Checlair17}. }

The climate model is  based on two equations describing  the time evolution of surface temperature  $T(\theta,t)$ and fraction of vegetation coverage $A(\theta,t)$, both as functions of time and the tidally locked latitude $\theta$ ( $\theta=0$ at the terminator and $\theta=\pi/2$ at the substellar point):
\begin{equation}\label{eq:T}
 \begin{aligned}
    {C_T}\frac{\partial T(\theta,t)}{\partial t} =&  \left\{ \left[ 1-\alpha(T,A) \right] S(a,L_\star,\theta) - R(T) \right\} + \\
    &{C}(T-\langle{T}\rangle )\\
 \end{aligned}
\end{equation}
\begin{equation}
\label{eq:A}
\begin{cases}
 \frac{\partial A(\theta,t)}{\partial t} =A \left[ \beta(T) (1 - A) - \gamma \right]  & \text{if $\theta\ge 0$}\\
A=0 & \text{if $\theta< 0$}
\end{cases}
\end{equation}
{ In equation (\ref{eq:T}), $C_T$ is the heat capacity, the first term on the right side represents the thermal energy balance between incoming stellar radiation $S$ and planetary outgoing flux $R$, the second term describes the zonal redistribution of heat by the atmospheric dynamics. Equation (\ref{eq:A}) is a logistic equation for the evolution of the fraction of land covered by vegetation where $\beta$ and $\gamma$ represent vegetation growth and death rates, respectively. Vegetation can affect the evolution of temperature by changing the total albedo $\alpha$. Since the {Trappist-1} planets are tidally locked and the nightside is never illuminated by the stellar radiation, we consider no vegetation growth in the planetary nightside where $A$ is fixed to $0$. The spatial average of the planetary temperature is computed as $\langle T\rangle= \int_{\theta_1}^{\theta_2} T(\theta) \cos\theta d\theta / \int_{\theta_1}^{\theta_2} \cos\theta d\theta$, where the extremes of integration ${\theta_1}$ and ${\theta_2}$ vary for global, dayside and nightside averages. \\
The stellar irradiation at the orbit of the planets, expressed as $S(a,L_\star,\theta)=({L_\star}/{4 \pi a^2} )\sin\theta$, depends on the star luminosity $L_\star$ and on the star-planet distance $a$. Since the planets are tidally locked, $S$ varies according to the geographical position on the day side, where $\sin \theta > 0$, and $S(a,L_\star,\theta) =0$ on the night side, where $\sin \theta < 0$. 
The planetary outgoing flux is $R(T) =  g(T) \sigma T^4$, where $g(T)$ is the grayness function that models the greenhouse effect  \citep{Sellers69}. The grayness function is written as  $g(T) =  {1 - m \tanh \left[ ({\langle T_{day}\rangle}/{T_0})^6 \right]}$, where $m$ parametrizes the atmospheric attenuation, $\langle T_{day}\rangle$ is the spatially averaged dayside temperature and  $T_0\sim 281$ K. This definition takes into account the infrared emission due to the planetary surface which is assumed to increase with the surface temperature and which implies a relaxation of the blackbody radiation hypothesis. \\ 
The total albedo $\alpha(T,A)$ depends on the fraction of ocean and land of the planetary surface as 
\begin{equation}
\alpha(T,A) = p(\theta) [\alpha_v A + \alpha_g (1-A) ] +  [ 1-p(\theta)] \alpha_o(T), 
\end{equation}
where $p(\theta)$ is the fraction of land covering the planetary surface as a function of the latitude, $\alpha_v$ and $\alpha_g$ are the albedos of vegetation and bare-ground, respectively. The albedo $\alpha_o$  of the ocean linearly depends on temperature as 
\begin{equation}
\alpha_o(T) = \alpha_{max} + ( \alpha_{min} - \alpha_{max} ) \left[ \frac{T - T_{low}}{T_{up} - T_{low}} \right] 
\label{alb_ocean}
\end{equation}
in the range of temperatures $T_{low} < T < T_{up}$; $\alpha_o=\alpha_{max}$ for an ocean completely covered by ice ($T \le T_{low}$) and  $\alpha_o=\alpha_{min}$ for an ice-free ocean ($T \ge T_{up}$).
The vegetation growth-rate is assumed to be a quadratic function of temperature \citep{Watson83,Alberti15,Rombouts15,Alberti18}
\begin{equation}
\beta(T) = max \left\{ 0; 1-k (T-T_{opt})^2\right\}
\end{equation}
with $k$ and $T_{opt}$ defining the temperature range inside which vegetation can grow, while the vegetation death-rate $\gamma$ is assumed to be constant \citep{Watson83,Alberti15,Rombouts15,Alberti18}.

The distance from the star and irradiation for the seven {Trappist-1} planets are taken from \citet{Gillon16}.
{All the planets are supposed to have an atmosphere with the Earth's composition and Earth-like greenhouse conditions ($m=0.6$)  have been considered. The fraction of planetary surface $p(\theta)$ occupied by land mimics that of the Northern hemisphere of a tidally locked Earth with the North pole at the terminator { and the substellar point  located at the Earth longitude 0$^{\circ}$}. The insolation shape and the fraction of planet surface occupied by land, used in the model, are shown in Figure \ref{insol}.}
\begin{figure}
\center
\captionsetup[subfigure]{labelformat=empty}
\subfloat[]{\includegraphics[scale=0.4]{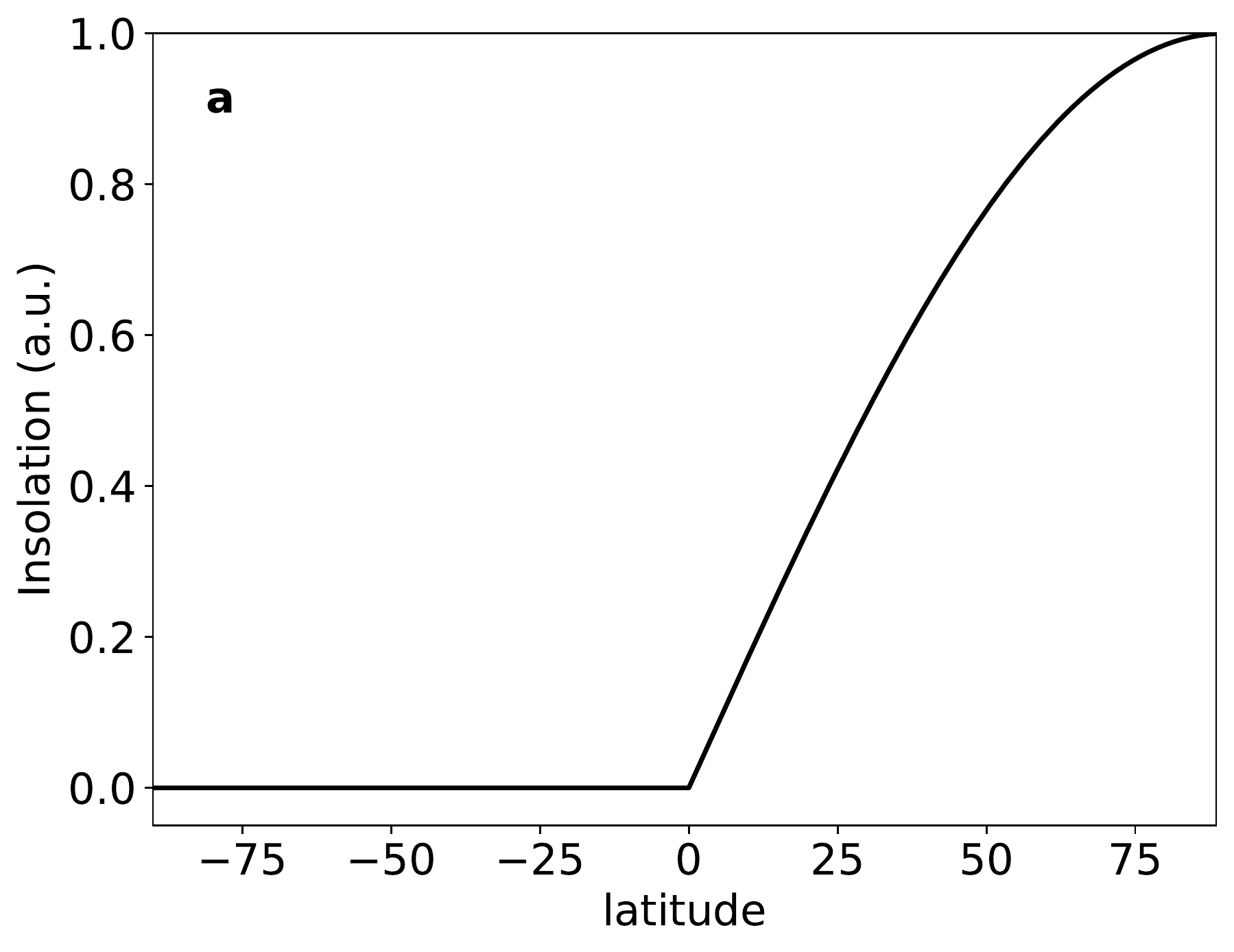}}
\subfloat[]{\includegraphics[scale=0.4]{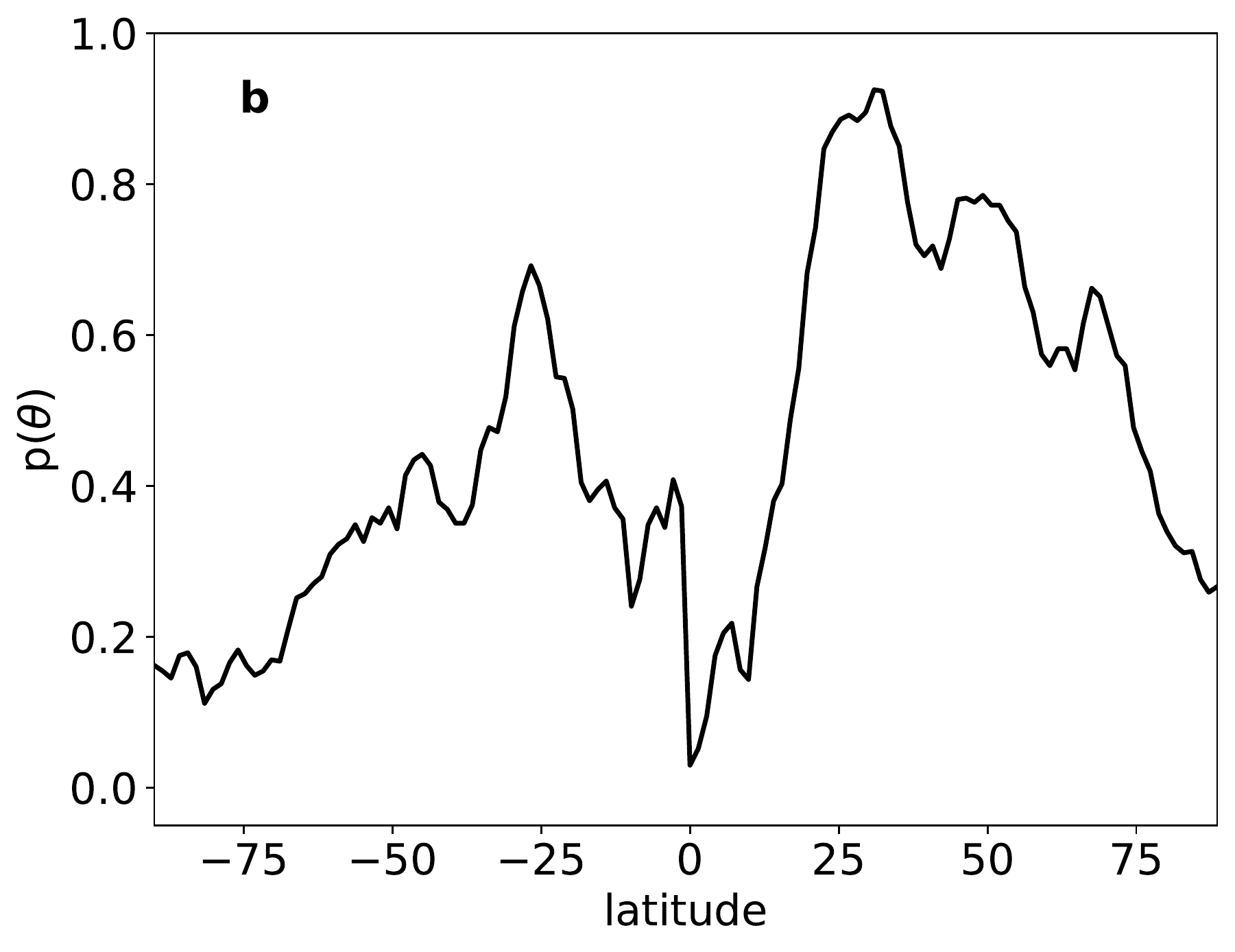}}
\caption{{ Insolation shape (a) and fraction of planet surface occupied by land (b) as functions of {tidally locked latitude.}} 
\label{insol}}
\end{figure}

In order to parametrize the atmosphere's large-scale heat transport through fluid motions, we introduce a  global relaxation term in the form  ${C}(T-\langle {T}\rangle )$  \citep[see e.g.][]{Checlair17}, where $\langle {T}\rangle$ is the spatially averaged temperature, that  mimics the processes determining heat redistribution from dayside to nightside on tidally locked planets. 
{The coupling coefficient ${C}$ is a constant calculated as} $\rho_s c_p C_D U$, where $\rho_s$ is the surface density of the atmosphere, determined by $\langle {T}\rangle$ and the surface pressure via the ideal gas law (a surface pressure of $p_s=1$ atm is assumed), $c_p$ is the atmospheric specific heat, $U$ is a typical surface wind speed and $C_D$ is a dimensionless drag coefficient having a typical value of $0.001$ over moderately rough surfaces \citep{Garrett94,Koll16}. A typical value of Earth's surface winds of $5$ m/s has been chosen \citep{Pierrehumbert11}.  

Equations (\ref{eq:T}) and (\ref{eq:A}) have been numerically integrated by using a second order explicit Runge-Kutta time scheme for the time marching and a latitude grid with a resolution $\sim 1.4^{\circ}$ ($128$ points). The initial latitudinal temperature profile is flat with a value equal to $T_{eq}$, determined by solving the radiative equilibrium for equation (\ref{eq:T}).  The initial vegetation coverage is randomly distributed over latitude. For all planets, the final temperature profiles are not affected by the initial condition for the vegetation distribution. Indeed, by varying the initial vegetation profile (e.g a constant vegetation coverage in the dayside or in the full planet) changes of the final temperature profiles are at most of the order of $2\%$. 

{Due to the uncertainty about the physical properties of the {Trappist-1} planets, we consider the parameters of the model, listed in Table \ref{tab1}, as for an Earth-like planet. The choice of the parameters presents some differences with respect to \citet{Alberti17} where a 0D EBM with the same equation for the fraction of vegetation coverage was developed. Here a $C_T$ value closer to the effective heat capacity of the Earth's ocean, of about $14\pm6$ W yr K$^{-1}$ m$^{-2}$ \citep{Schwartz07}, has been chosen. At the same time lower values of $\alpha_{min}$ and $\alpha_{max}$ for the ocean albedo has been considered here. Indeed, the stellar spectrum of the M-dwarf like Trappist-1 is significantly redder than a Sun like star and at longer wavelengths the albedos of ice-free and ice-covered ocean are lower than with respect to what expected for the Earth \citep{Shields13}. Finally, a temperature $T_{low} = 253$ K, below which the ocean is completely covered by ice, and a value of $k$ allowing the growth of vegetation up to $\sim 317$ K have been chosen. All the other parameters are the same as in \citet{Alberti17}.

 }
\begin{deluxetable}{ccc}
\label{tab1}
\tablecaption{Definition and values of the parameters of the model.}
\tablewidth{0pt}
\tablehead{
\colhead{Symbol} & \colhead{Value} &\colhead{Units}  \\
}
\startdata
$C_T$  & $15.2$ & W yr K$^{-1}$ m$^{-2}$ 	\\
$\alpha_v$ & $0.1$ &  $\cdot\cdot\cdot$				\\	
$\alpha_g$ & $0.4$ & 	$\cdot\cdot\cdot$			\\
$\alpha_{max}$	& $0.55$ & 	$\cdot\cdot\cdot$			\\
$\alpha_{min}$	& $0.2$ & 	$\cdot\cdot\cdot$			\\
$T_{low}$	 & $253$ & K				\\
$T_{up}$ & $300$ & K				\\
$T_{opt}$ 	 & $283$ &  K				\\
$k$ & $0.0008$ &  yr$^{-1}$ K$^{-2}$	\\
$\gamma$ & $0.1$ & yr$^{-1}$			\\
$c_p$ & $10^{-3}$ & J K$^{-1}$ K$^{-1}$ \\
$m$ & $0.6$ & $\cdot\cdot\cdot$\\
\enddata
\end{deluxetable}
\section{Results}
\subsection{The effect of vegetation}
To investigate the role of vegetation in self-regulating planetary temperatures, two cases have been considered: a planet with no vegetation, only covered by bare soil and oceans that can be frozen, and a planet covered by vegetation in addition to soil and sea. Temperature and vegetation profiles at some time steps of the simulation for all the planets of the {Trappist-1} system are shown in  Figure \ref{multi}  for both cases with and without vegetation. 
\begin{figure}
\center
\includegraphics[scale=0.4]{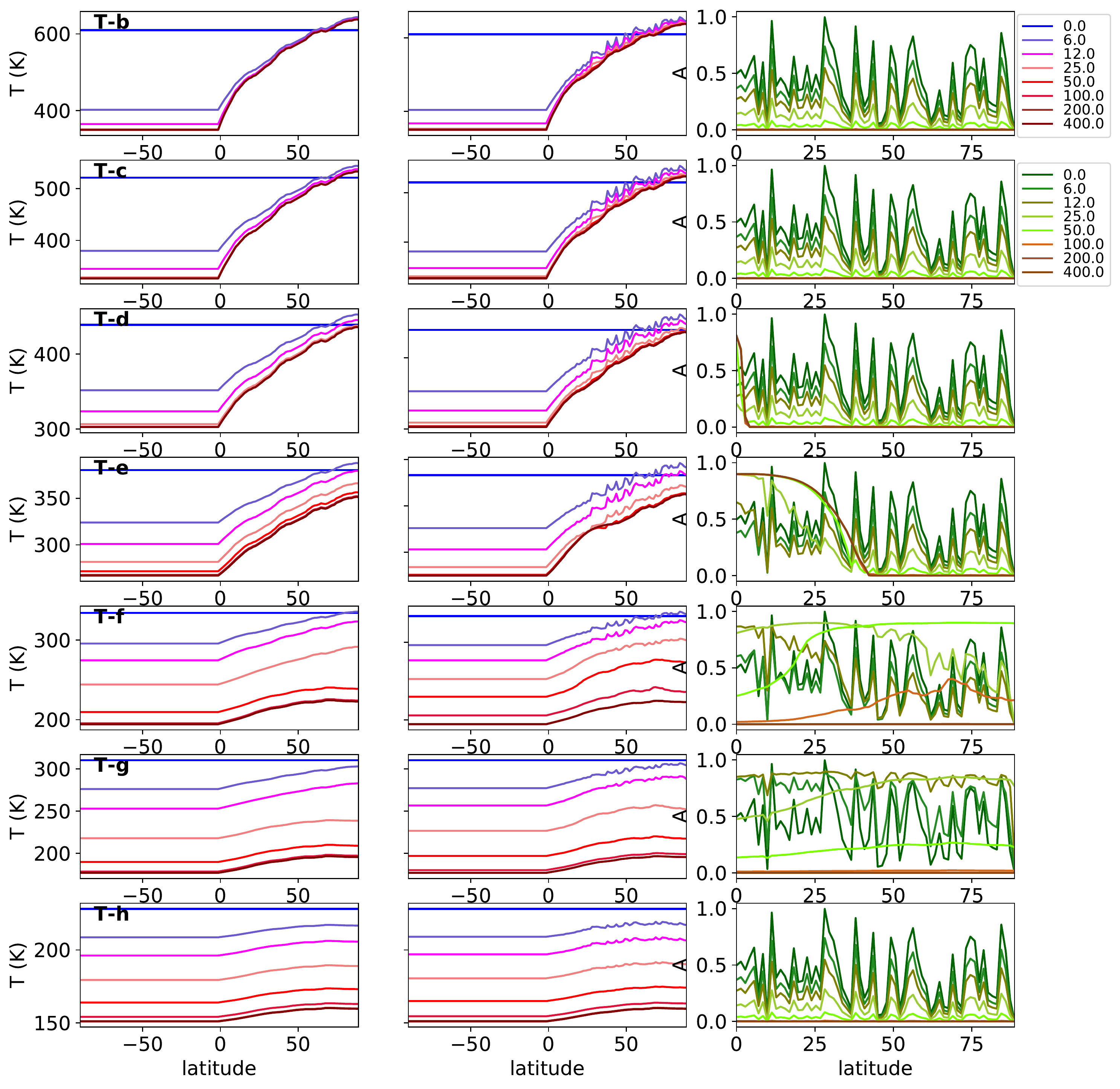}
\caption{{Temperature profiles for the cases without (left column) and with vegetation (central column) and { the fraction of vegetation coverage $A$
 (right column)} at different time steps of the simulation. Each row corresponds to a different planet of the {Trappist-1} system. Different colors correspond to different times of the simulation. Times in the legends are in years.} 
\label{multi}}
\end{figure}
The average dayside, nightside, and global temperature resulting at the last step of the simulation  are shown in Table \ref{tab2}. { Figure \ref{T_veg_noveg} shows the final (at the last step of the simulation) planetary surface temperature as a function of the latitude.} \\
\begin{deluxetable}{cccc}
\label{tab2}
\tablecaption{Nightside T$_n$, dayside T$_d$ and average T$_{avg}$ temperatures for the seven {Trappist-1} planets in the Earth-like scenario. Last line refers to T-e for the case with vegetation coverage.}
\tablewidth{0pt}
\tablehead{
\colhead{Planet} & \colhead{T$_n$ (K)} &\colhead{T$_d$ (K)} & \colhead{T$_{avg}$ (K)} \\
}
\startdata
T-b&$350$& $515$& $433$ \\
T-c&$327$& $438$& $382$\\
T-d&$303$& $370$& $336$\\
T-e&$267$& $307$& $287$\\
T-f&$194$& $211$& $203$\\
T-g&$177$& $188$& $182$\\
T-h&$151$& $156$& $153$\\
\hline
T-e&$275$& $320$& $297$\\
\enddata
\end{deluxetable}
When comparing the results for the cases with and without vegetation, significant differences { in the temperature profiles} are observed only for T-e. For this planet, indeed, the surface temperature at some latitudes allows vegetation to grow, which in turn has a feedback on the temperature. Vegetation growth is also allowed for T-d in a very restricted range of tidally locked latitudes (about $4^\circ$) near the terminator (third row of Figure \ref{multi}). Since occurring only in a small fraction of the planetary surface, this growth does not significantly affect the T-d temperature profile, which is the same as the case without vegetation (the difference in the average temperature between the cases with and without vegetation is 0.01 K).
For the other planets the vegetation growth rate is almost zero and, since the death rate is constant in the model, they evolve toward a state with no vegetation at all. 
 Since the planets are tidally locked, so that the stellar irradiance is $0$ in the dark side, the temperature is constant in the nightside of all the planets. 
\begin{figure}
\center
\includegraphics[scale=0.5]{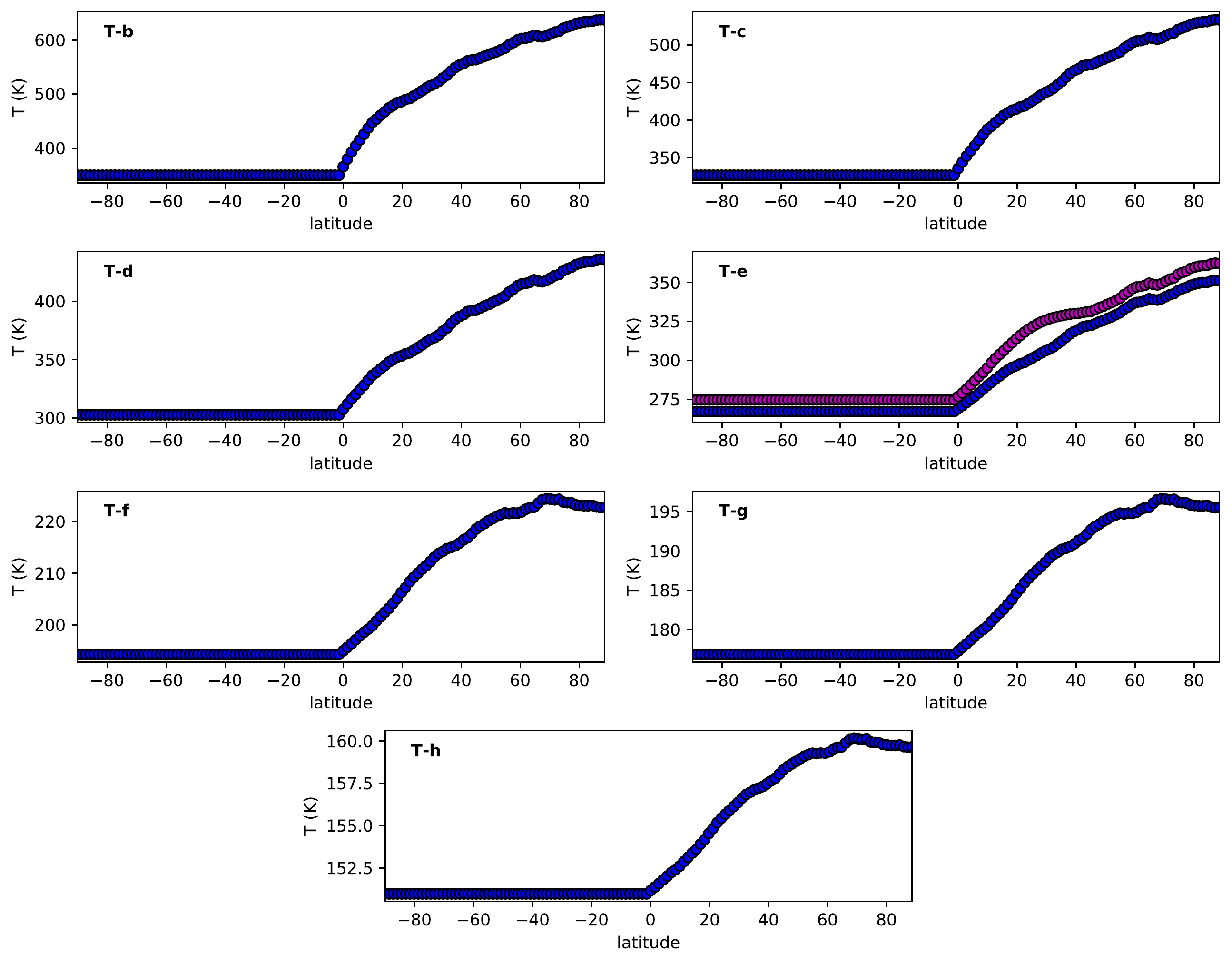}
\caption{{Surface temperature as a function of the latitude at the last step of the simulation for all the {Trappist-1} planets. Results for a model with (magenta curves) and without (blue curves) vegetations are shown.} 
\label{T_veg_noveg}}
\end{figure}
Inner planets (b, c, and d), show dayside temperature of 515, 438, and 370~K, higher or very close to boiling point of water. This indicates that large fractions of the daysides of these planets, the places more suitable to develop life since receiving direct stellar radiation, do not sustain liquid water at the surface and they are likely hot and dry. These results agree with those obtained by more complex 3D climate models \citep{Gillon16, Wolf17}. On the other hand, for external planets (f, g, h) the star irradiance is  too low to sustain surface temperatures above the melting point of water and these planets should be in a snowball state \citep{Gillon16, Wolf17}. 
These considerations can be quantified by introducing a measure of the habitable fraction (HF) of the planet as the percentage of the planet's dayside surface residing in the HZ. HF quantifies if and where the planet can sustain liquid water on its surface: by assuming a CO$_2$-H$_2$O-N$_2$ atmosphere and under the assumption of an Earth-like geology for the resulting greenhouse effect and carbon-silicate weathering cycle, this condition corresponds to have a local surface temperature in the range $273$-$373$ K. For the three inner planets the habitable fraction is at most $38\%$ (being $3\%$, $11\%$ and $38\%$ respectively) and for external planets it is $0\%$.  \\
{For inner planets and T-e the temperature profile shows a quite monotonic trend increasing toward high $\theta$.  The external planets show, instead, a temperature decrease for $\theta \gtrsim 65^\circ$ due to the steep decrease in $p(\theta)$ beyond the relative maximum at $\theta \sim 65^\circ$.}
This behavior is due to the effect of the ocean on the planetary albedo. For the hotter planets (b, c, d, e), the temperature profile { is monotonic like} the insolation pattern: the contribution of the stellar irradiance is high and the albedo fluctuations due to the bare soil and ocean (the ocean albedo $\alpha_0$ has the minimum value $\alpha_{min}$) introduce small variations. On the other hand, for external colder planets, the stellar irradiance is lower and the latitudinal pattern of the ocean albedo (which is higher, being $\alpha_0 = \alpha_{max}$ since the planet is completely covered by ice), determined by the function $p(\theta)$, {contributes more in defining the general behaviour of the temperature profile.}

For the case with no vegetation, planet e represents the best candidate for a habitable world showing an average global temperature of $287$~K. T-e dayside temperatures, always below {the boiling point of water}, allow to sustain liquid water and make the planet resident in the habitable zone; on the other hand, the nightside, with an average temperature of $267$~K, is ice-covered. The habitable fraction of T-e is very high, reaching the $93\%$. 
When a vegetation coverage, with an albedo of $0.1$, is introduced in the model, the surface temperature profile of T-e changes. An increase of the T-e dayside surface temperature at low-intermediate latitudes, forming a bump in the latitudinal temperature profile, is observed. This is located is the range of latitudes where the conditions for vegetation growth are favorable, thus affecting the resulting surface temperature. In the dayside the net effect of vegetation is to produce a micro-climate, in a given latitude range, where the surface temperature locally changes. As a consequence, the habitable fraction of T-e increases and reaches ${100}\%$.  Since for Earth-like conditions T-e is the best candidate to habitability, in the following we will focus on this planet. 
\subsection{Changing vegetation types.}
One of the peculiarities of the model is the possibility to explore the vegetation-albedo feedback acting together with the planet's ice-albedo feedback. Since the albedo depends on the vegetation species, we investigate how the change of vegetation type can affect the planetary temperature. To this purpose the values of the vegetation albedo $\alpha_v$ in the model are changed to take into account the effect of different types of vegetation. To cover the range of albedos associated with different species \citep[from $\sim 0.08$ for a conifer site to $\sim 0.2$ for green grass,][]{Betts97} we  evaluated the T-e temperature for different $\alpha_v$ namely $0.05, 0.1, 0.15$ and $0.2$ (figure \ref{veg_type}). The global shape of the temperature profile does not change by varying the vegetation albedo but the dayside/total average temperatures decrease from $321$/$298$ K to $318$/$295$ when $\alpha_v$ increases from $0.05$ to $0.2$.  The changes in the temperature profile are observed, as expected, in the bump at intermediate latitudes while everywhere else the temperature profiles collapse on almost the same values. 
\begin{figure}
\center
\includegraphics[scale=0.5]{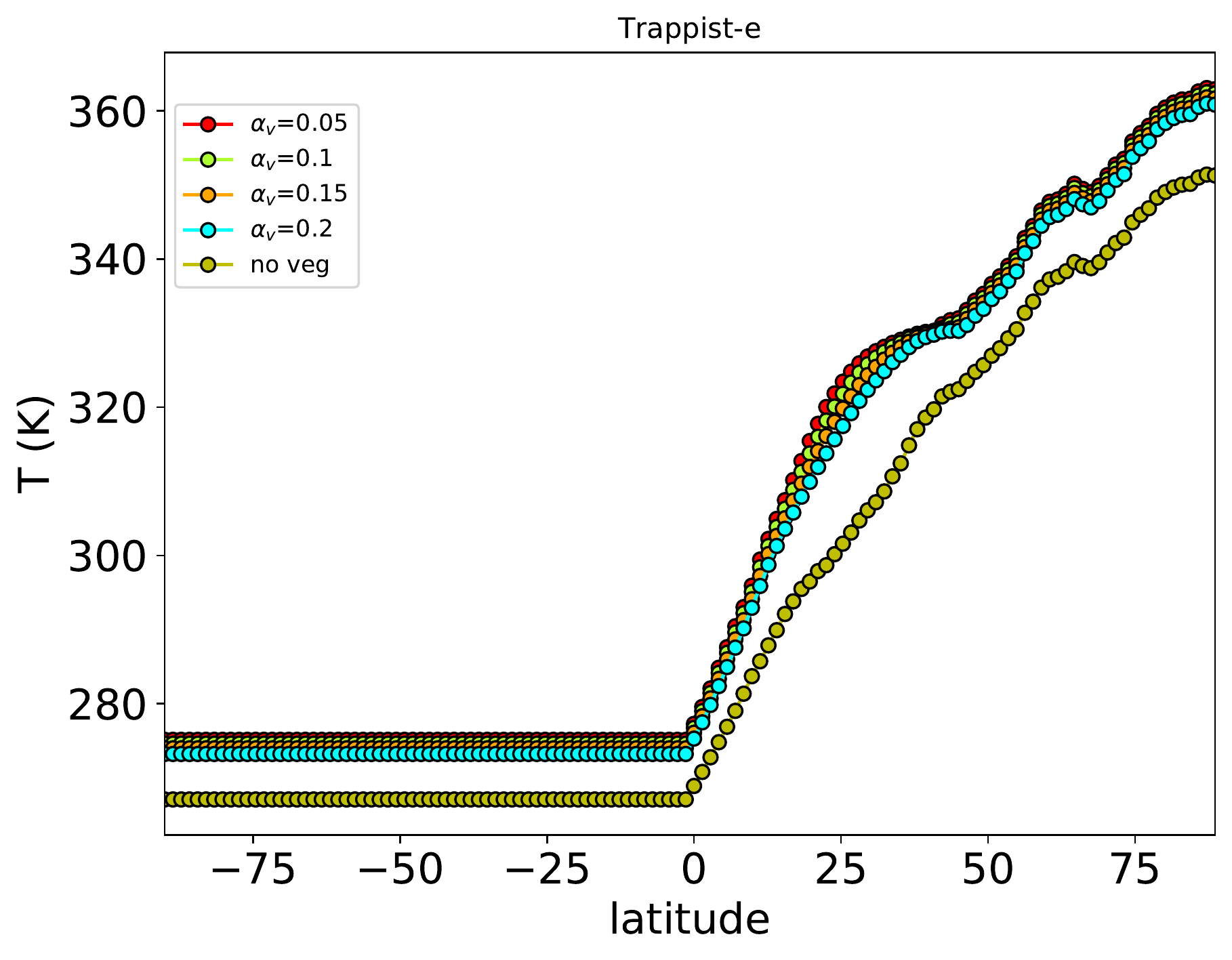}
\caption{{ Surface temperature for T-e, as a function of the latitude at the last step of the simulation, obtained by using different vegetation types characterized by different albedos.} 
\label{veg_type}}
\end{figure}
\subsection{Changing the land-sea distribution}
An important role in defining the planetary temperature profile is played by the land-sea distribution, since it can deeply modify the planetary albedo. The planetary response to changes in land-sea percentage can be very complex. This is because the total albedo is affected also by the vegetation coverage that can change if land coverage is increasing or decreasing. 
To investigate the effect of changes in the land-sea distribution on the resulting T-e planetary temperature, three extreme cases have been considered: i) $99\%$ of the nightside is covered by land (with the dayside completely covered by sea; in this case there is no vegetation growth); ii)  $99\%$ and iii) $50\%$ of the dayside are covered by land (with the nightside completely covered by sea). Resulting temperature profiles for the three different land-sea distributions are shown in Figure \ref{land_distrib} together with profiles for the Earth-like case.  
Case i) shows higher temperatures when compared with results of cases ii) and iii) with no vegetation. This is because at these temperatures the total albedo for case i) is just the ocean albedo which favors higher temperatures having a lower value with respect to the albedo of the bare soil with no vegetation. {The behavior of the temperature profiles for cases ii) and iii) without vegetation at high $\theta$ , where the albedo is only determined by the percentage of soil and ocean, reflects the behavior observed at the same latitudes for the planets including vegetation}.
{When the growth of vegetation is taken into account,  temperature profiles for  ii) and iii)  confirm that the vegetation coverage plays a role in structuring the temperature of the planet. Indeed they show, as for the Earth-like case, a bump at low latitudes with increased width for iii) due to a broader range of latitudes at which vegetation can grow.
For the three temperature profiles higher values of temperature are observed at high latitudes where the albedo of i) reaches the minimum value $\alpha_{min}$  and  the total albedo of ii) and iii) also includes the contribution of the bare soil only (due to the high temperatures vegetation does not grow), thus resulting in higher temperatures.  A difference of about $10$ K is observed between ii), for which the total dayside albedo is almost completely determined by the bare soil, and iii), for which the albedo is reduced due to a higher ocean percentage. \\
On the other hand, at lower latitudes (between $0^{\circ}-18^{\circ}$) the increase of the albedo of the ocean, {being the temperature between $T_{low}$ and $T_{up}$ (see equation (\ref{alb_ocean})), and the growth of vegetation, producing a decrease of the total albedo, result in comparable temperatures for the three cases. At these latitudes, case ii) shows a larger increase of temperature than iii) since a larger} land percentage implies a higher fractions of vegetation coverage and higher decrease of the total albedo.
}

For the case with no vegetation, a larger day-night temperature difference is observed when a lower percentage of land coverage (case iii) is considered. The gradient is larger for the case i) when the land coverage is confined to the night side and decreases for iii) and ii), respectively.  This result agrees with several studies showing that dry planets produce larger day-night temperature differences with respect to moist ones due to a decrease of the efficiency of the  heat transport \citep[e.g.][]{yang2013}. On the other hand, for planets with vegetation but different land-sea coverage, the day-night temperature difference is almost the same. This suggests that the vegetation coverage acts to counteract, by varying the planetary albedo, the heat redistribution along the planet.
\begin{figure}
\center
{\includegraphics[scale=0.5]{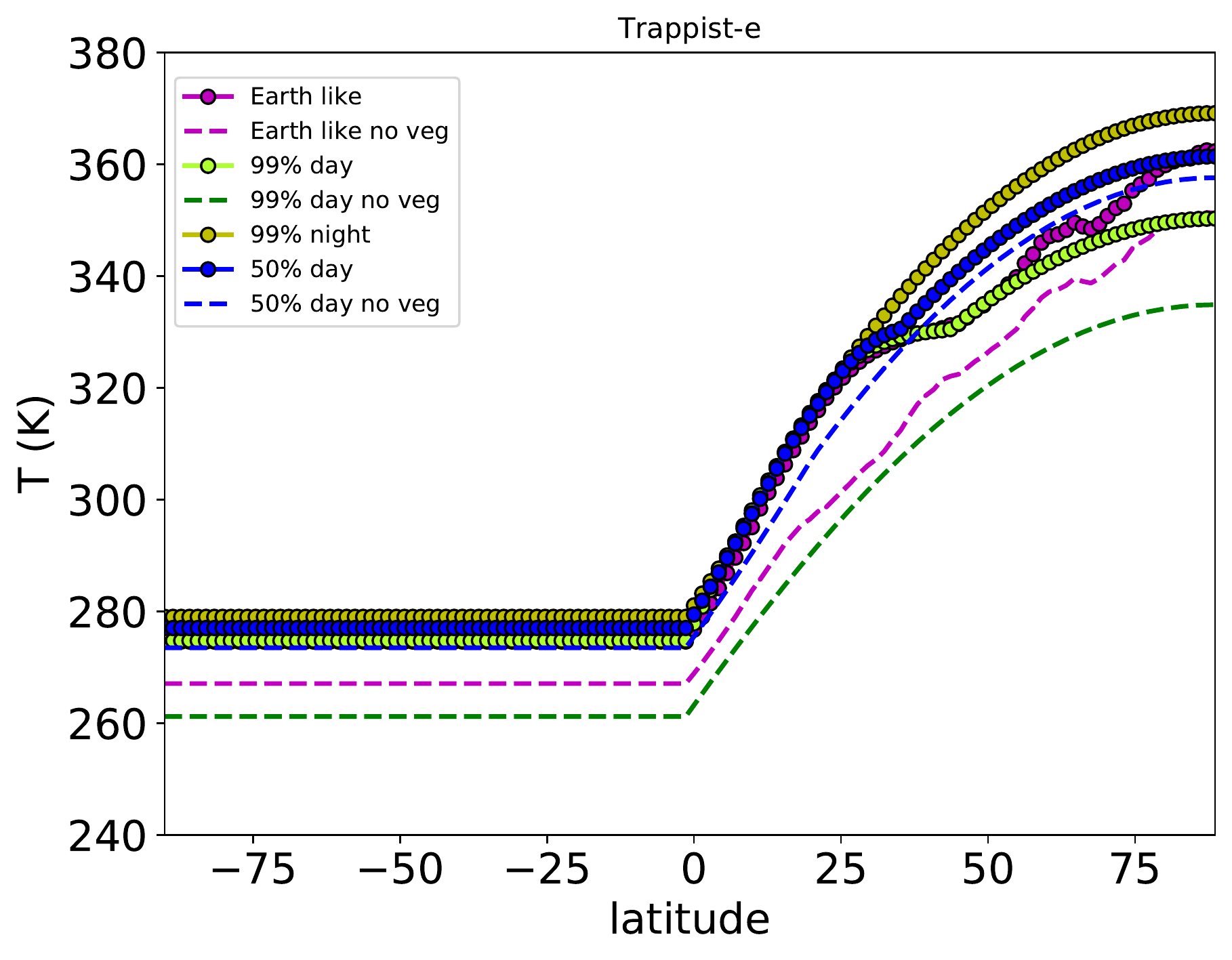}}
\caption{{ Surface temperature for T-e, as a function of the latitude at the last step of the simulation, obtained by using different land-sea profiles and the same initial random distribution of vegetation.} 
\label{land_distrib}}
\end{figure}
\subsection{Changing the composition of the atmosphere}
At the last stage of our study, we investigated the response of the system when the atmospheric composition is changed { by increasing the $CO_2$ content of the Earth-like T-e atmosphere}. Changes in the $CO_2$ content act on the large-scale heat transport in equation (\ref{eq:T}) and also affect the greenhouse effect by changing the atmospheric attenuation $m$ in the grayness function. According to \citet{Sellers69}, we can roughly estimate that a variation of a factor $10$ in the $CO_2$ content changes $m$ by about $10\%$. {This means that for an increase of the $CO_2$ content of a factor $5$, $10$, $15$, with respect to an Earth like atmosphere, the parameter $m$ increases to $0.63$, $0.66$, $0.69$, accordingly. 
The results of the planetary temperatures for different atmospheric compositions are shown in Figure \ref{atm_comp}.
As expected, the increase of the $CO_2$ content in the atmosphere produces a higher average planetary temperature: an increase of $\sim 18$~K in the average planetary temperature, with respect to an Earth-like planet, is found when $CO_2$ content is increased by a factor $15$.} The addition of greenhouse gas in the atmosphere mainly affects the areas with no vegetation. The differences among temperature profiles obtained with different $CO_2$ contents are {lower} in the range of latitudes where vegetation can grow while they are enhanced at {higher} latitudes where only bare soil and ocean are present. {For a more quantitative comparison, we evaluated the average temperature for the curves of Figure \ref{atm_comp} in the ranges of latitude $\Delta\theta_l=[0^{\circ},24^{\circ}]$, where vegetation grow, and $\Delta\theta_h=[50^{\circ},180^{\circ}]$, where there is no vegetation. Average temperatures calculated in both $\Delta\theta_l$ and $\Delta\theta_h$ increase, with respect to the Earth-like atmosphere, when the $CO_2$ content is increased. However, the increase of temperature in $\Delta\theta_l$ are lower by 1, 3, 6 K with respect to those calculated in $\Delta\theta_h$, when the  $CO_2$ content increases by factors $5$, $10$, $15$. \\ 
This indicates that even when the greenhouse effect is enhanced to higher values with respect to the Earth, thus producing a global increase of the planetary temperature, the lowering of the albedo due to the vegetation growth is still effective in structuring the temperature profiles.} The allowance of vegetation growth in the model, by locally changing the albedo, can thus reduce the effects of the greenhouse effect with respect to the same model with no vegetation.
\begin{figure}
\center
{\includegraphics[scale=0.5]{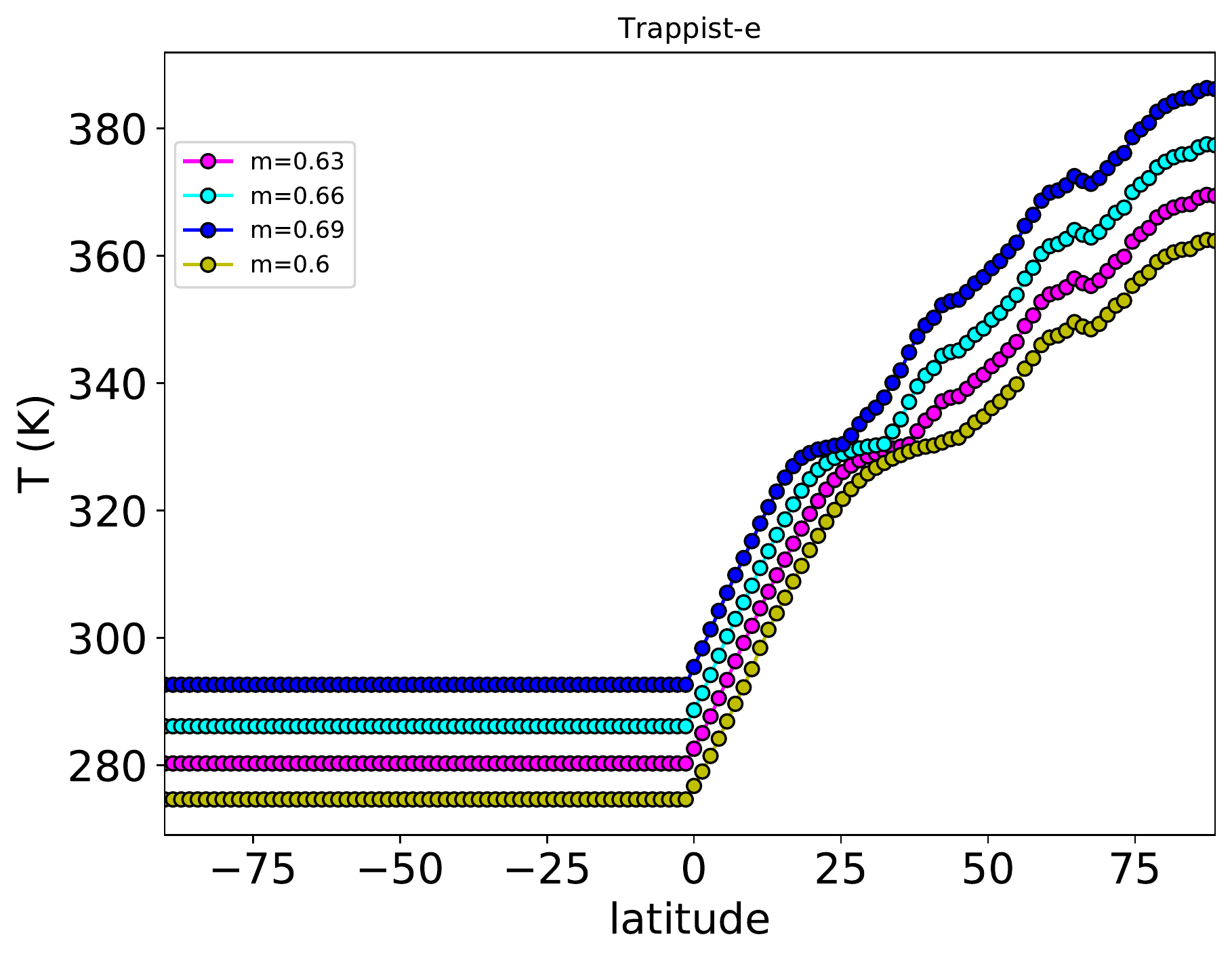}}
\caption{{ Surface temperature as a function of the latitude at the last step of the simulation for T-e obtained by using different $CO_2$ fractions in the atmosphere. {$m=0.6$: Earth like atmosphere.  $m=0.63$, $0.66$, $0.69$: $CO_2$ content increased by $5$, $10$, $15$ times with respect to a Earth like atmosphere.}} 
\label{atm_comp}}
\end{figure}
\section{Conclusions}
A 1-D EBM model, with latitudinal dependency, was developed and used to investigate the vegetation feedback on the climate of the planets in the {Trappist-1} system. Besides the vegetation-albedo feedback, the model includes also a parametric description of the greenhouse effect and the possibility to vary the land-sea distribution. Firstly, the temperature profiles of all the Trappist-1 planets were obtained by using model parameters of an Earth-like planet. However, since for a M-dwarf like Trappist-1 the stellar spectrum is redder than for the Sun, the maximum and minimum values of the ocean albedo have been lowered with respect to the Earth-like case. {T-d and T-e are the only planets for which vegetation is able to grow at some latitudes. Since for planet d vegetations grows only in a very small latitude range close to the terminator,  the temperature profile is not affected and it is the same as the case without vegetation. On the other hand, vegetation growth occurs for T-e  in a larger fraction of the planetary surface giving rise to significant changes in the temperature profiles. \\
Since characterized by an average global temperature of 287/297 K, for the cases without/with vegetation, T-e presents the most favorable conditions for habitability, in agreement with the results of previous works  \citep{Wolf17,Bolmont17,Bourrier17,Kopparapu17};  actually the presence of a vegetation coverage sightly increases the habitable fraction of the planet. {These findings differ from results of \citet{Alberti17} who found, by implementing a 0D EBM model with vegetation coverage, that the inner planets T-b, T-c, and T-d are the best candidates for habitability while planet T-e can present water oceans only for greenhouse effect conditions different from those of Earth. These differences can arise both from differences in the model equations, such as the addition of a spatial dimension to the model and the inclusion of a day-night heat transport term in the 1D model described here, and differences in the values of some parameters (see Section \ref{sec:model}).}

By varying  the vegetation albedo we investigated the effect of different types of vegetation on the planetary temperatures. Changes in vegetation albedo (between $0.05$ and $0.2$, in the range of values measured for the different species commonly observed on the Earth) produce quite small changes in the  temperatures and {habitable surface fraction}, the latter changing of at most ${0.1\%}$.

We then studied how changes in the land-sea distribution affect planetary temperatures. When no vegetation growth is considered, a planet with the $99\%$ of ocean in the dayside reaches the highest temperature with respect to planets with other land-sea distributions.
For planets with the nightside completely covered by sea but with different percentages of land in the dayside, $99\%$ and $50\%$, the final temperature profiles show a similar shape  of the Earth-like case. Nevertheless, different land-sea distributions result in broader/wider range of latitude where the vegetation can grow and enhanced temperature difference between low and high latitudes. 

A significant effect is produced  by changes in the atmospheric composition by adding different quantities of $CO_2$ (which both affects the large-scale heat transport and increases the grayness parameter) to an Earth's like planetary atmosphere. {It is found that the increase of $CO_2$ leads, as expected, to higher average planetary temperatures: for example, when the $CO_2$ content of the planetary atmosphere is increased by a factor 15, with respect to an Earth like atmosphere, an increase of about 18~K in the average planetary temperature is found. However, the increase of greenhouse gas abundance in the atmosphere mainly affects the regions with no vegetation. The effect of vegetation coverage in settling the final planetary temperature profile is still relevant even for greenhouse effect much higher than the Earth like case.}

A new set of parameters for the {Trappist-1} planets, slightly differing from those provided by  \citep{Gillon16}, have been published by \citep{Delrez18}. New planet-star distance are on average increased of about the $2\%$ while irradiations decrease by $7\%$ on average. The results of our model remain substantially unchanged if these new parameters are used. Indeed, due to the decrease of the stellar irradiation, the global average temperature decreases and the habitability slightly changes ($326/291$ K and $44\%/97\%$ for planet d and e respectively, for the case with vegetation).

\acknowledgments
We acknowledge the anonymous reviewer for the useful comments.

\bibliographystyle{aasjournal} 
\bibliography{references} 

\end{document}